\documentclass[A4,aps,prl,reprint,superscriptaddress,nofootinbib]{revtex4-1}
\usepackage{graphicx}
\usepackage{dcolumn}
\usepackage{bm}
\usepackage{epstopdf}
\usepackage{amsmath}
\usepackage{hyperref}
\usepackage{color}
\usepackage{cancel}
\usepackage{ulem}

\def\fun#1#2{\lower3.6pt\vbox{\baselineskip0pt\lineskip.9pt
  \ialign{$\mathsurround=0pt#1\hfil##\hfil$\crcr#2\crcr\sim\crcr}}}

\begin{document}


\title {Probing $\Xi N$ interaction through inversion of spin-doublets in $\Xi N \alpha \alpha$ nuclei}

\author{E.\ Hiyama}
\affiliation{Department of Physics, Tohoku University, Sendai, Japan,
980-8578}
\affiliation{Nishina Center for Accelerator-Based Science, RIKEN, Wako,
351-0198,  Japan}

\author{M. Isaka}
\affiliation{Sicence Research Center, Hosei University, Tokyo 102-8160,Japan}

\author{T. Doi}
\affiliation{Interdisciplinary Theoretical and Mathematical 
Sciences Program (iTHEMS), RIKEN, Wako 351-0198, Japan}
\author{T. Hatsuda}
\affiliation{Interdisciplinary Theoretical and Mathematical 
Sciences Program (iTHEMS), RIKEN, Wako 351-0198, Japan}

\date{\today}
%
\begin{abstract}
A new way to study the spin-isospin dependence of the $\Xi N$ interaction is explored
 through the energy levels of  $\Xi N\alpha$ and $\Xi N\alpha \alpha$ systems with $\alpha$ 
  being a spectator to attract  the $ \Xi N$ pair without changing its spin-isospin structure.
 By using the Gaussian expansion method (GEM)   with the state-of-the-art $\Xi N$ potential obtained from lattice 
  QCD calculations,    {it} is found that $\Xi N\alpha \alpha$ has  
   spin-doublet bound states with $J^{\pi}={1^-}$ and  $2^-$
    in {both} isospin triplet and singlet channels.  The inversion of the $1^-$-$2^-$  spin-doublet between the iso-triplet and the iso-singlet
   is found to be strongly correlated with the relative strengths of the $\Xi N$ interaction in the 
    $^{11}{\rm S}_0, ^{13}{\rm S}_1,^{31}{\rm S}_0$ and $^{33}{\rm S}_1$ channels.
      The $(K^-, K^+)$ and $(K^-,K^0)$ reactions on  the  $^{10}$B  target are proposed 
 to produce those bound states.
\end{abstract}

\maketitle


\noindent
{\it Introduction.$-$} 
In recent years,  considerable progress has been made in studying the 
  unknown hyperon interactions in the $S=-2$ channel, especially the $\Xi N$
interaction.
Experimentally, a bound $^{15}_{\Xi}{\rm C} (\Xi+  ^{14}{\rm N})$ hypernucleus 
was observed using the emulsion detector \cite{Nakazawa,Nakazawa2018,Hayakawa,Yoshimoto},
which indicates that the $\Xi$-nucleus potential is attractive
and the $\Xi N$-$\Lambda\Lambda$ coupling is weak.
 Also, the femtoscopic analysis of the  $\Xi N$ momentum correlation in 
$p$-Pb and $p$-$p$ collisions by ALICE Collaboration at LHC \cite{LHC1,LHC2} shows
 that the spin-isospin averaged $\Xi N$ interaction is attractive at low energies.

 Experimental studies so far have limited access to 
the spin-isospin decomposition of the  $\Xi N$ interaction, and 
 it is important to explore theoretically possible ways to make   the decomposition. 
   In Ref. \cite{Hiyama2008}, 
it was pointed out, by using the Gaussian expansion method (GEM)
 with  the  Nijmegen $\Xi N$ potentials,
that   the $^7_{\Xi}{\rm H}(\Xi + ^6{\rm He})$ and
$^{10}_{\Xi}{\rm Li}(\Xi + ^9{\rm Be})$ hypernuclei may have bound states
  and are suited to extract the information on the spin-isospin independent part of the $\Xi N$
interaction.  
More recently,  in Ref.\cite{Hiyama2020}, 
 the binding energies of $\Xi NN$ and $\Xi NNN$ hypernuclei were examined by GEM and
  the first principles lattice QCD interaction (the HAL QCD potential)  \cite{Sasaki2020}, where it is found 
  {that} a shallow bound state {exists} with $T=0, J^{\pi}=1^+$ in the $\Xi NNN$ system due to the moderately large 
 attraction of the HAL QCD potential in  the $\Xi N ( ^{11}{\rm S}_0)$ channel.\footnote{
  Here,    we employ the spectroscopic notation $^{2T+1,2s+1}{\rm S}_J$
    to classify the S-wave $\Xi N$  interaction where $T$, $s$, and $J$ stand for total isospin, total spin, and total angular momentum, respectively.}
Subsequently,  several bound states for light $\Xi$ hypernuclei ($A=4,5$ and 7) were suggested
 by using  the no-core shell model \cite{chiral2021} with  a possible strong attraction   
 of the chiral effective field theory interaction in the $\Xi N ({^{33}{\rm S}_1})$ channel.
 
  The purpose of the present paper is to explore a
 robust and unambiguous way to extract the spin-isospin component
 ($^{11}{\rm S}_0, ^{13}{\rm S}_1,^{31}{\rm S}_0$ and $^{33}{\rm S}_1$)
 of the  $\Xi N$ interaction by considering 
 the systems, $\Xi N \alpha$ and  $\Xi N \alpha \alpha$:
 Since $\alpha$ is a spin-isospin saturated system,  the  $\Xi N$ interaction is directly
  linked to  the spin-isospin structure of these systems.
   In particular,  we calculate their binding energies within the framework of
 three- and four-body cluster models using the HAL QCD $\Xi N$ potential, and 
  propose possible experiments to produce such states through the $K^-$ induced reactions.

\

\noindent
{\it Few-body method.$-$} 
 {For $\Xi N\alpha$ and $\Xi N\alpha \alpha $,}
the total Hamiltonian is
given by
\begin{equation}
H=K+\sum_{a,b}V_{ab} +V_{\rm Pauli},
\label{eq:hami}
\end{equation}
where $K$ is the kinetic-energy operator,
$V_{ab}$ is the interaction between the constituent particle
$a$ and $b$, and $V_{\rm Pauli}$ is the Pauli
projection operator to be defined below.

\begin{figure}[htb]
\begin{center}
\vspace{-1cm}
\includegraphics[scale=0.4]{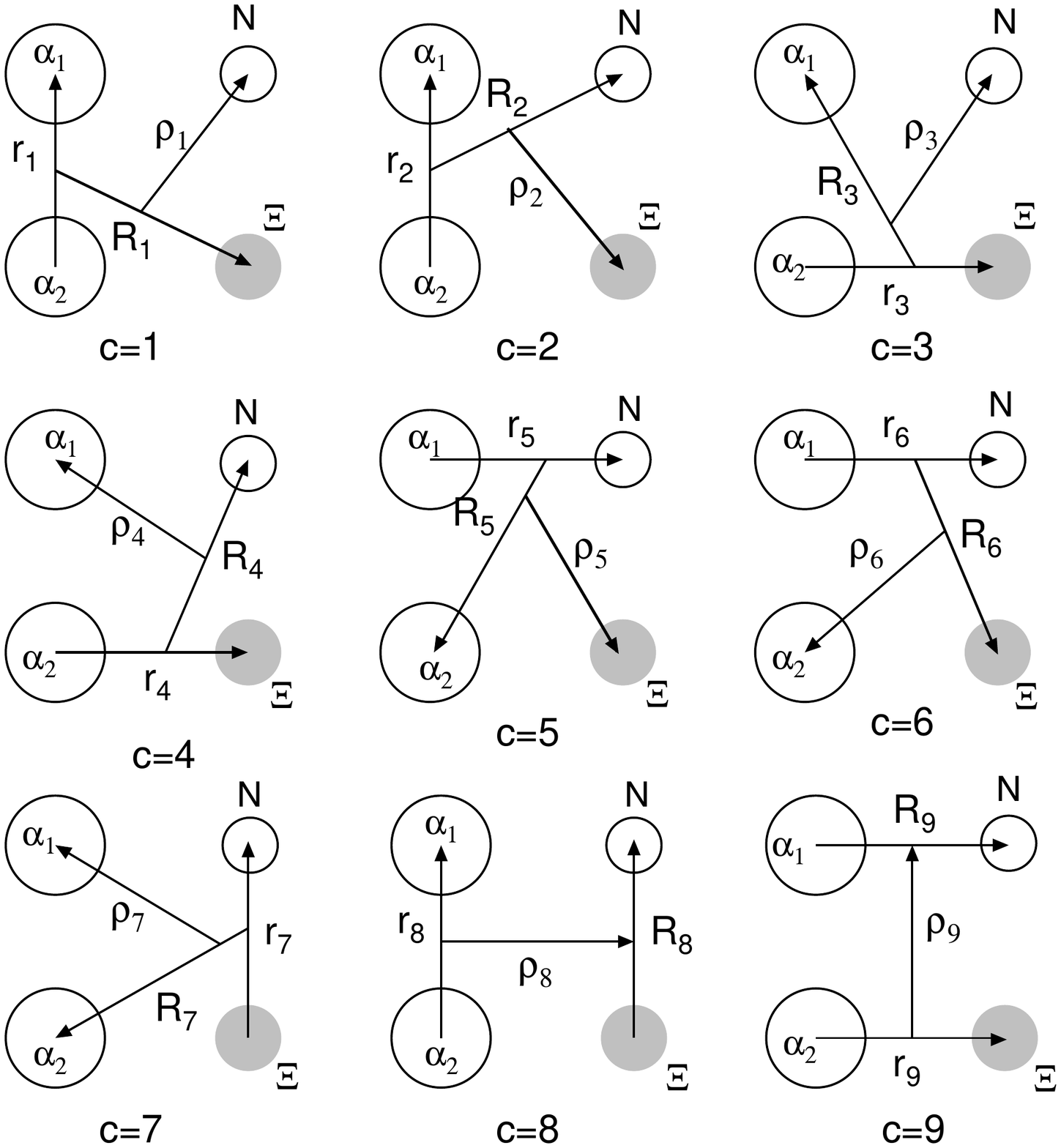}
\vspace{-2.2cm}
\caption{Jacobi coordinates for the nine channels to describe the $\Xi N\alpha \alpha $ four-body system.
 }
\label{fig:a10}
\end{center}
\end{figure}

The total wavefunctions of  $\Xi$ hypernuclei with $A=6,10$ 
  satisfying the Schr\"{o}dinger equation, $(H-E)\Psi_{JM,TT_z}^{(A=6,10)}=0$,
are described as  sums of amplitudes
of all rearrangement channels {with the $\alpha$ cluster(s)  (1$-3$ (1$-$9) channels for $A= 6 (10)$)}
in the $LS$ coupling scheme. Shown in Fig.1 is the case for $A=10$
 with the wave function,
\begin{eqnarray}
\Psi_{JM,TT_z}^{(A=10)} &=& \sum^9_{c=1} \sum_{n,N,\nu}
\sum_{\ell, L, \lambda} \sum_{S,I,K}
C^{c}_{n\ell NL\nu \lambda SIK}  \nonumber \\
&\times & {\cal S}_{\alpha \alpha} {\Big[}\Phi(\alpha_1)\Phi(\alpha_2)
[\chi_\frac{1}{2}(N)\chi_\frac{1}{2}(\Xi) ]_S \nonumber \\
&\times & \big[[\phi^{(c)}_{n\ell}({\bf r_c})
\psi_{NL}^{(c)}({\bf R_c}) \big]_I 
\xi_{\nu \lambda}^{(c)}(\rho_c)\big]_K {\Big]}_{JM} \nonumber \\
&\times& [\eta_\frac{1}{2}(\Xi)\eta_\frac{1}{2}(N)]_{T T_z}.
\label{eq:a10}
\end{eqnarray}
Here ${\cal S}_{\alpha \alpha}$ stands for the symmetrization operator for
exchange of two $\alpha$ clusters.
$\chi_\frac{1}{2}(N)$ and $\chi_\frac{1}{2}(\Xi)$ are the spin wavefunction of the 
nucleon and $\Xi$ particle, respectively. 
$\eta_\frac{1}{2}(N)$ and $\eta_\frac{1}{2}(\Xi)$ are the isospin wavefunction of the 
nucleon and $\Xi$, respectively. 
Throughout this paper, we use the isospin-averaged nucleon mass and that of the $\Xi$ mass.
 The Coulomb interaction is fully taken into account.  The mixing of the wave functions between the $T=1$ and $T=0$ 
 states is the second-order effect in isospin symmetry breaking and is not considered.

Following GEM \cite{Hiyama2003}, we take the functional forms,
$\phi_{n\ell m}({\bf r})=r^{\ell} e^{{-(r/r_n)}^2}Y_{\ell m}(\hat{\bf r})$, 
 $\psi_{NLM}({\bf R}) = R^L  e^{{-(R/R_N)}^2}Y_{LM}(\hat{\bf R})$, 
 and $\xi_{\nu \lambda \mu}({\rho}) = \rho^\lambda  e^{{-(\rho/\rho_\nu)}^2}Y_{\lambda \nu}(\hat{\bf \rho})$,
where the Gaussian range parameters are chosen to follow the 
geometrical progressions:
$r_n =r_1 u^{n-1}\ (n=1-n_{\rm max})$,
 $R_N = R_1 v^{N-1}\  (N=1-N_{\rm max})$, 
 and $\rho_\nu = \rho_1 w^{\nu -1} \  (\nu =1-\nu_{\rm max})$. 
 The eigenenergy $E$
and  the coefficient $C$ in  Eq.(\ref{eq:a10})
are to be determined by the Rayleigh-Ritz
variational method.

As for $V_{\alpha \alpha}$ and  $V_{N \alpha}$,
we employ the potentials which reproduce reasonably well 
the low-energy $\alpha \alpha$ and $N \alpha $ scattering phase shifts \cite{Hasegawa,Kanada}.
{An in-medium fudge factor, 0.955,  is multiplied to $V_{N \alpha }$ when it is used in the 
 systems containing  $\alpha\alpha$, with the factor determined from the binding energy (1.57 MeV) 
  of $^9{\rm Be}$.} The Coulomb potentials involving $\alpha$  cluster(s)
are constructed by folding the proton distribution in the $\alpha$ cluster.

The Pauli principle for the nucleons belonging to $\alpha$ and $j (={N},\alpha)$
is taken into account by the orthogonality
condition model (OCM) \cite{OCM} where
 the Pauli projection operator in Eq.(\ref{eq:hami}) is given by
\begin{equation}
V_{\rm Pauli}= \gamma \sum_f \enskip|\phi_f({\bf r}_{j\alpha}) \rangle \langle \phi_f({\bf r}'_{j \alpha})| \enskip.
\end{equation}
 The Pauli-forbidden relative wavefunction between $\alpha$ and $j$
 is denoted by   $\phi_f ({\bf r}_{j \alpha})$,
 where  $f=0{\rm S}$ for $j={N}$ and  { $f=0{\rm S},1{\rm S} ,0{\rm D}$  for $j=\alpha$ are {chosen} according to the standard
OCM procedure.}
The Gaussian range parameter $b$ of the
single-particle $0s$ orbit in the
$\alpha$ particle ($(0s)^4$)
is taken to be $b=1.358$ fm  as in
the literature \cite{Motoba}.
In the actual calculation, the strength $\gamma$  is taken
to be $10^4$ MeV, which is
large enough to push the unphysical
forbidden state to high energies
while keeping the physical states unchanged.
 
 \
 
\noindent
{\it $\Xi N$ potential.$-$} 
As for the $\Xi N$ interaction, we employ the results 
 based on  {the isospin symmetric}  (2+1)-flavor 
lattice QCD simulations in a spacetime volume $L^4 = (8.1 \ {\rm fm})^4$
 with a lattice spacing $a$=0.0846 fm and at the nearly physical quark masses  $(m_{\pi}, m_K)$=(146, 525) MeV  \cite{Sasaki2020}.
 HAL QCD Collaboration derived the $\Xi N$-$\Lambda \Lambda$ coupled-channel potentials
from the simulation data   at the Euclidean time $t/a=11,12,13$. 
         As discussed in Ref.\cite{Hiyama2020},   
      the lattice QCD data of the coupled-channel $\Xi N$-$\Lambda \Lambda$ system
   are  fitted  with multiple  Gaussians and a Yukawa  form with $(m_{\pi}, m_K) = (146, 525)$ MeV,
    and the results are  extrapolated to the isospin symmetric physical point  $(m_{\pi}, m_K) = (138, 496)$ MeV. 
    In the $^{11}{\rm S}_0$ channel,   an effective single-channel $\Xi N$ potential is introduced by 
     renormalizing the coupling to $\Lambda\Lambda$   into  
   a single range Gaussian whose parameter is chosen to reproduce the $\Xi N$ phase shift obtained by the channel coupling.
   
     Key properties of the resulting potentials are  that
    the $^{11}$S$_0$ channel is most attractive,
  $^{13}$S$_1$ and $^{33}$S$_1$ channels are weakly attractive,  and
  the $^{31}$S$_0$ channel is weakly repulsive (See Fig.2(a)).
  In Ref.\cite{Hiyama2020}, the lightest bound $\Xi$ hypernucleus was found to be $\Xi NNN$ system
   using this potential.
   {We use this single-channel $\Xi N$ potential throughout this paper for all partial waves,
      which is valid for the long-range part of the interaction relevant to the present study.}
In the following,  central values of the observables such as the binding energy  are obtained from the data at $t/a=12$
  with systematic errors  estimated from those at  $t/a=11$ and  13.  Uncertainties associated with the statistical errors 
  of lattice QCD data are comparable to these systematic errors.

\

\noindent
{\it $\Xi \alpha$ potential.$-$} 
The $\Xi \alpha $ potential is obtained by folding 
 the above $\Xi N$ interaction 
based on the $(0s)^4$ configuration of $\alpha$. 
Since the isospin-spin averaged $\Xi N$ potential reads $V_0=[V( ^{11}{\rm S}_{0})+3 V(^{13}{\rm S}_{1})
+3 V(^{31}{\rm S}_{0})+ 9 V(^{33}{\rm S}_{1}) ]/16$,
 the $\Xi$-$\alpha$ interaction is dominated by the $^{33}{\rm S}_1$ channel.
 The attraction of the HAL QCD potential from lattice QCD
  in this  channel is considerably weaker 
  than those of the Nijmegen ESC08c potential  (Fig.1 of \cite{Hiyama2020}) 
  and of the chiral NLO potential in \cite{chiral2021}.
  Thus, we find that the  binding energies $B_{\Xi\alpha}$ {with respect to $^{5}_{\Xi}$H} with the Coulomb interaction are as small as 
$0.64, 0.45, 0.25$ {MeV} for   $t/a=11,12,13$, respectively, {with the width $\Gamma \sim 5 $ keV},
 so that $\Xi\alpha$ is likely to be the Coulomb assisted bound state.
 In contrast,  $B_{\Xi\alpha}$ is about 2.16 MeV  in the chiral NLO potential \cite{chiral2021}
 where the {attraction in the $ ^{33}{\rm S}_1$ channel}  is assumed to be large.

\

\noindent
{\it  $\Xi N\alpha $ system.$-$} 
Let us first discuss the  $\Xi N\alpha $ system with $T=1$
and $T=0$.  Since  the ground state of $^5$He has
 $J^\pi = 3/2^-$, total isospin-spin states of possible $\Xi N\alpha $ nuclei
 are $(T,J^\pi)=(1,1^-), (1,2^-), (0,1^-)$, and $ (0,2^-)$,  where 
  $J^\pi=2^- (1^-) $ corresponds  to the
 spin-parallel (anti-parallel)  $\Xi N$ pair.
   Thus, $\Xi N$ interactions in  $^{31}{\rm S}_0$,
  $^{33}{\rm S}_1$, $^{11}{\rm S}_0$, and $^{13}{\rm S}_1$  channels are {primarily (but not entirely)}  related to  
  $\Xi N\alpha $ systems with $(1,1^-)$, $(1,2^-)$, $(0,1^-)$, and $(0,2^-)$, respectively, 
  {as  summarized in Table \ref{tab:MRC}.}
In the HAL QCD potential, {$^{31}{\rm S}_0$} is repulsive 
and $^{33}{\rm S}_1$ is only weakly attractive (Fig,2(a)), so that
  $T=1$ $\Xi N\alpha $ bound states do not appear in  $1^-$ and  $2^-$ 
  states.
It is unlikely to find bound states,  unless the strength of the potential in 
  the  $^{33}{\rm S}_1$ channel is artificially increased by a factor of  2. 
In $T=0$,
there {generally} arise no bound states too in both  $1^-$ and  $2^-$ states, since
the attractions in $^{11}{\rm S}_0$ and {$^{13}{\rm S}_1$} channels are not 
large enough.
 (Only when we use the HAL QCD potential at $t/a=11$,
  there remains a possibility of  a very shallow bound state in $1^-$ state.)
Experimentally, one may try to   produce $\Xi N\alpha $ with $T=1$ and $T=0$
 by the   $(K^-, K^+)$ and $(K^-,K^0)$ reactions on the $^6$Li target, respectively.
However, it would be difficult to find
bound states in A=6 systems according to the $\Xi N$ potential based on lattice QCD.

\begin{figure*}[htb]
\vspace{-3.5cm}
\begin{tabular}{cc}
\begin{minipage}{0.45\hsize}
\vspace{-1.5cm}
\centering
\includegraphics[scale=0.38]{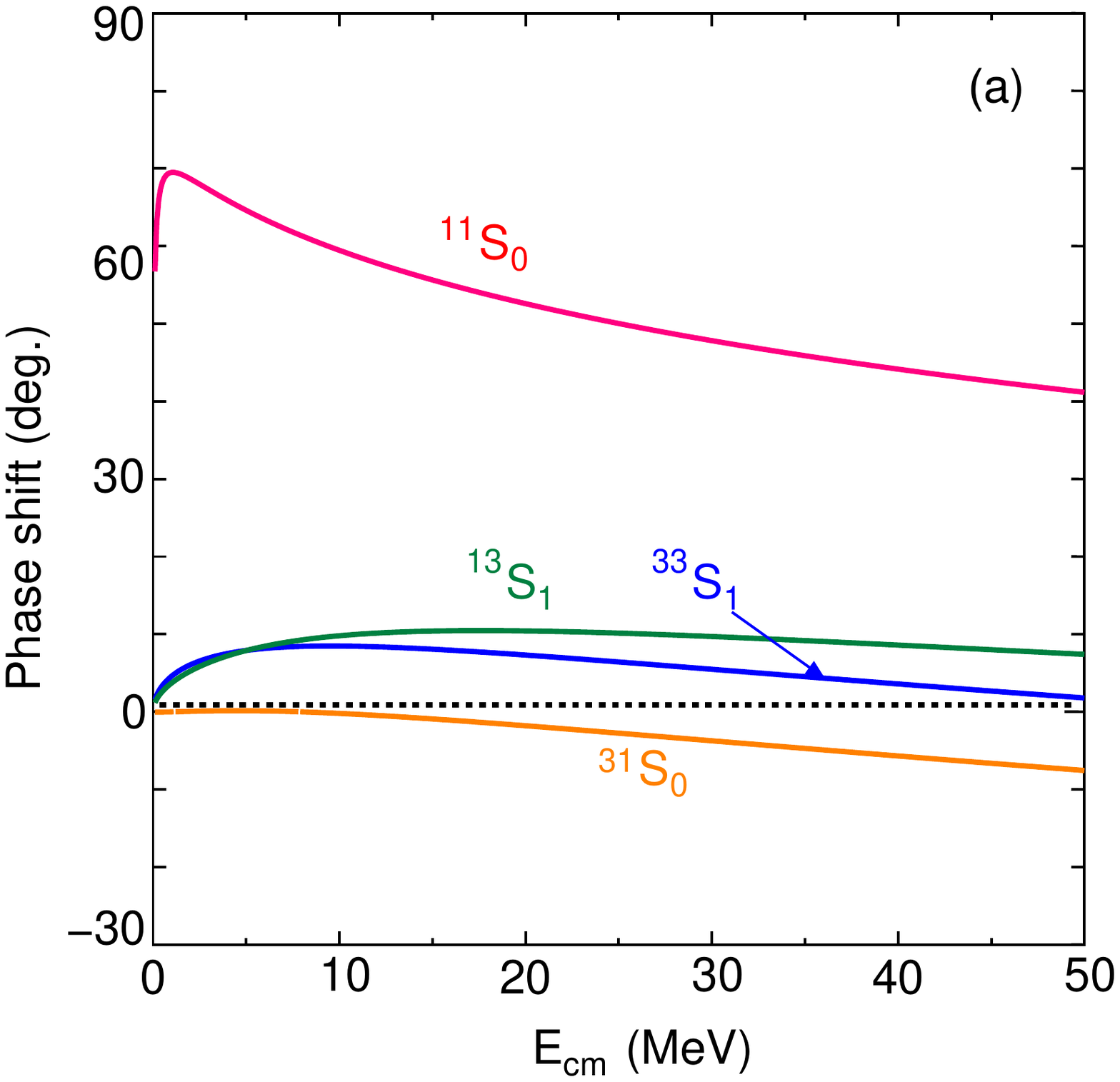}
\end{minipage} &
\begin{minipage}[h]{0.47\hsize}
\centering
\includegraphics[scale=0.45]{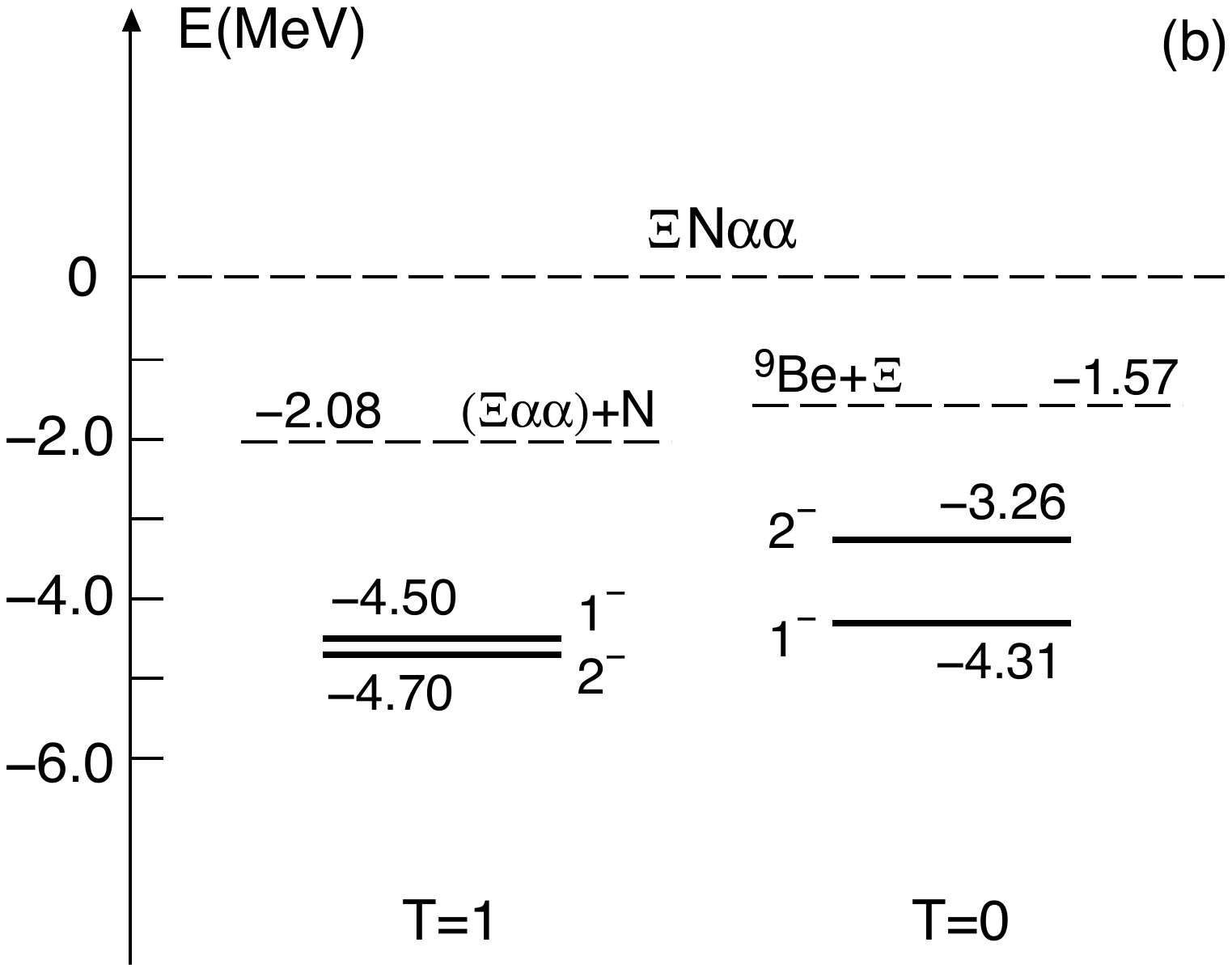}
\end{minipage}
\end{tabular}
\vspace{-4cm}
\caption{(a) $\Xi N$ phase shifts obtained from the HAL QCD potential.
 The figure is adapted and modified from \cite{Sasaki2020}.
(b) The calculated energy level of $\Xi N\alpha \alpha $ 
system. The energy is measured with respect to {$\Xi N \alpha \alpha $} four-body
breakup threshold. }
\label{fig:be10xlevel}
\end{figure*}

\

  \begin{table}[t]
\caption{{Possible spin-doublets in $\Xi N \alpha$ and $\Xi N \alpha \alpha$}  together with the $\Xi N$ channel primarily related to each state.}
\label{tab:MRC} 
\vskip 0.3cm 
  \begin{tabular}{|c|c|c|c|c|}
\hline
$(T, J^{\pi}$)   &  $ (1,1^-) $   & $(1,2^-)$    & $ (0,1^-) $   & $(0,2^-)$    \\
 \hline \hline
primary $\Xi N$ channel     &  $^{31}{\rm S}_0$ &  $^{33}{\rm S}_1$ & $^{11}{\rm S}_0$ & $^{13}{\rm S}_1$  \\
  \hline
\end{tabular}
\end{table}

\noindent
{\it  $\Xi N\alpha \alpha$ system.$-$}
 Let us now study possible four-body bound states by adding an
extra $\alpha$  to $\Xi N \alpha $.
Such a four-body system has  three-body bound states;
$^9{\rm Be} (T=1/2, J^\pi = 3/2^-)$ with the experimental binding energy of 1.57 MeV, and 
$\Xi^-  \alpha \alpha (T=1/2, J^\pi = 1/2^+)$ with the theoretical binding energy of  $ 2.08^{+0.77}_{-0.63}$ MeV.
 The first number is  reproduced by the three-body calculation by GEM,  while the second number is obtained by
GEM with HAL QCD potential.
Then possible $\Xi N\alpha \alpha$ nuclei with $T=1, 0$
 would have $J^{\pi}=2^-(1^-)$ corresponding to the spin-parallel
(anti-parallel)  $\Xi N$ pair.  Accordingly,   the $\Xi N$ channels primarily  contributing to these  states
are those given    in Table \ref{tab:MRC}.

{By considering  the  state $|\Xi^- n \alpha \alpha \rangle $ for $T=1$ and 
   $|\Xi^0 n \alpha \alpha  + \Xi^- p \alpha \alpha \rangle$ for $T=0$ by 
   the four-body calculation with GEM  and  the HAL QCD potential.
    we found the bound states with 
 $(T, J^\pi)=(1, 1^-), (1, 2^-), (1,3^- )$  and $(0, 1^-), (0, 2^-)$.   
 Summarized in Table \ref{tab:energy}  are the energy levels $E$  relative to the four-body breakup threshold  $\Xi$+$N$+$\alpha$+$\alpha$
 with systematic errors. 
  Note that the states in Table \ref{tab:energy} are located  below the lowest two-body breakup
    thresholds, $(\Xi\alpha\alpha)+N$ for $T=1$ and $^9{\rm Be}+{\Xi}$ for $T=0$.  
      Also shown in Table \ref{tab:energy}  are  the widths $\Gamma$ estimated perturbatively by using the HAL QCD  $\Xi N$-$\Lambda\Lambda$
  coupling potential. }
 
 \begin{table}[t]
\caption{
Energy levels of $\Xi N\alpha \alpha $ for
$T=1$ and 0 with the HAL QCD potential.  $E$ 
 is measured relative to the $\Xi+ N+\alpha+\alpha$ four-body breakup threshold.
$\Gamma$ is
   the decay width through the process $\Xi N \rightarrow \Lambda \Lambda$.
   Central values are 
    evaluated by the HAL QCD potential at $t/a=12$ and the systematic errors
    are estimated by the potential at $t/a=11$ and 12. } 
 \label{tab:energy} 
\vskip 0.3cm 
\begin{tabular}{|c|c|ccc|}
\hline
 &$J^\pi$ &$1^-$   & $2^-$    &$3^-$  \\
 \hline \hline
  $T=1$           &$E$ (MeV)   &  $-4.50_{-0.80}^{+1.04}$ &$-4.70_{-0.83}^{+1.09}$  
 &$-2.47^{+1.09}_{-0.84}$  \\ 
    &$\Gamma$  (MeV)  &  $\ \ 0.02^{+0.01}_{-0.01}$  & $\ \ 0.02^{+0.01}_{-0.00}$ 
  &$\ \  0.02^{+0.01}_{-0.00}$  \\
\hline
    $T=0$   &$E$ (MeV)     &$-4.31^{+1.28}_{-1.04}$  &$-3.26^{+1.10}_{-0.90}$ 
&-   \\
               & $\Gamma$ (MeV)  &$\ \ 0.04^{+0.01}_{-0.01}$  &$\ \  0.03^{+0.00}_{-0.01}$   
&-   \\
\hline
\end{tabular}
\end{table}

\begin{figure*}[hbt]
\vspace{-2.5cm}
\includegraphics[scale=0.40]{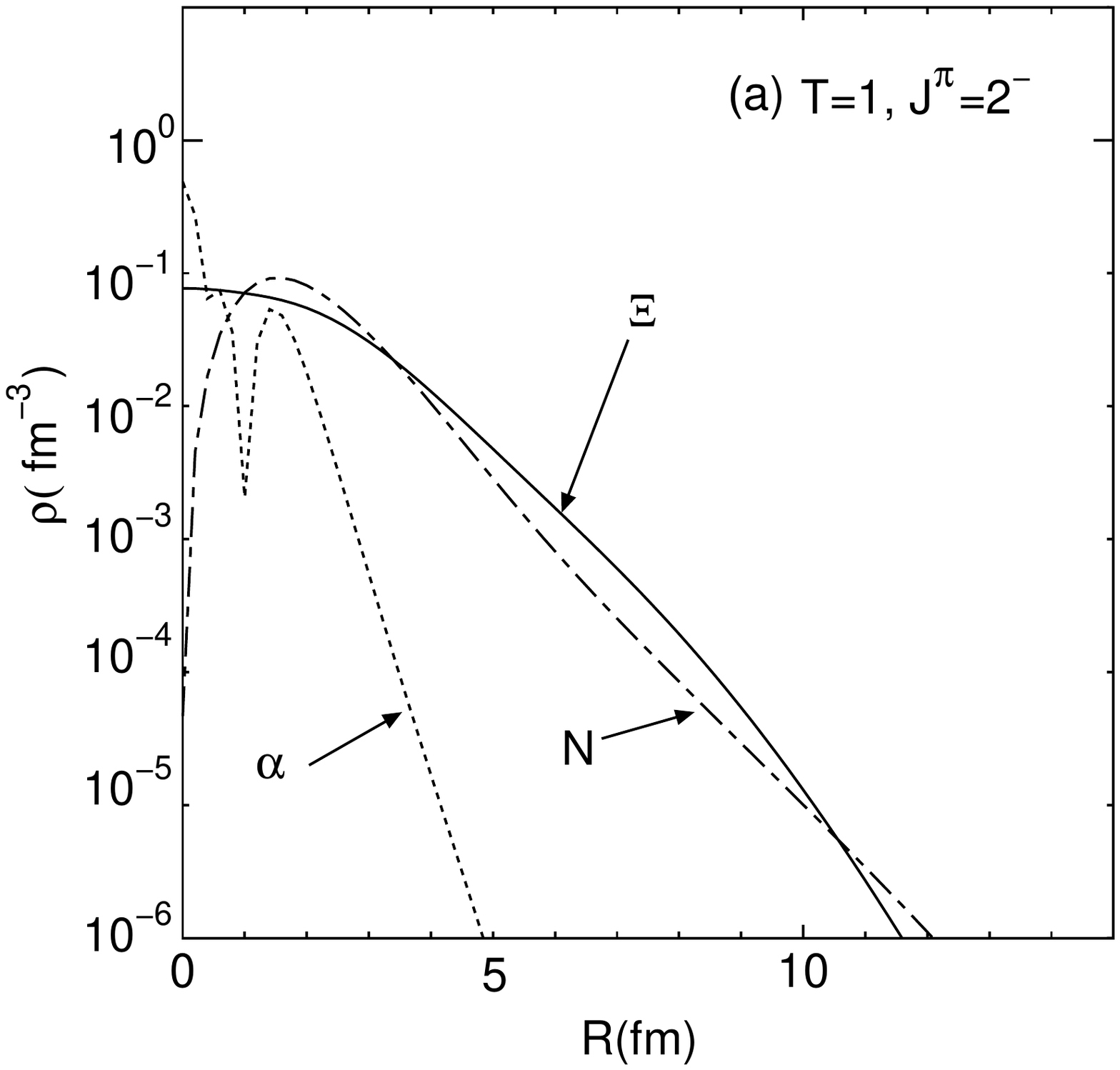}
\includegraphics[scale=0.40]{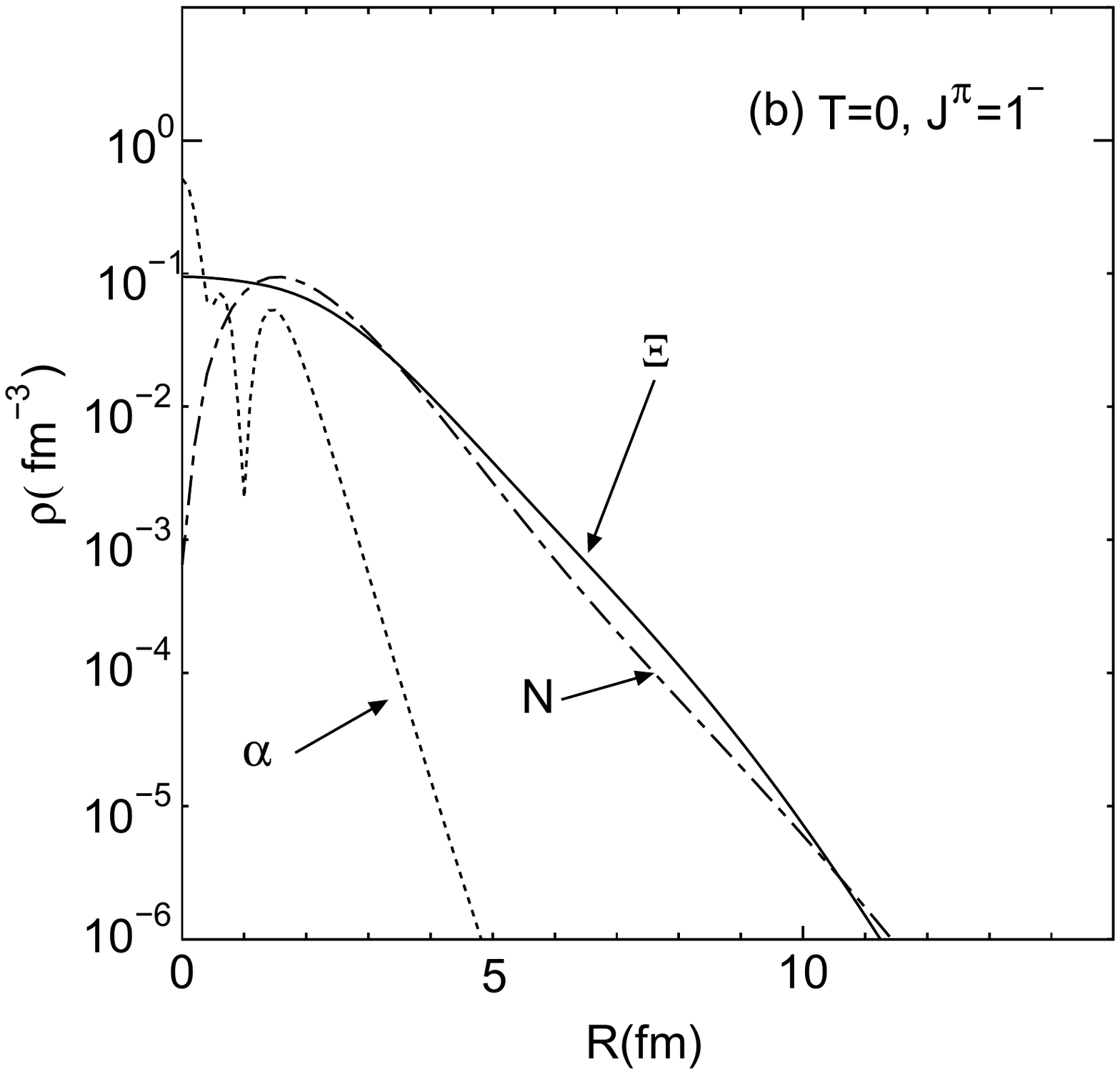}
\vspace{-2.5cm}
\caption{{Density distributions of $\alpha, \Xi$ and the valence nucleon $N$ with respect to the center of the mass of $\alpha$-$\alpha$
    in the ground states of the $\Xi N \alpha \alpha$ system with $T=1$ (a) and with $T=0$ (b).}}
\end{figure*}

\

\noindent
{\it  Inversion of spin-doublets.$-$} 
{ In Fig.2(b), we show   the  level structure of the lowest spin-doublets, $1^-$-$2^-$,  for $t/a=12$.}
For $T=1$,  the ground (1st excited) state has {$2^- (1^-)$  
 in which dominant $\Xi N$ interaction is in the $^{33}{\rm S}_1$ ($^{31}{\rm S}_0$)} channel
  being very weakly attractive (weakly repulsive) in the HAL QCD potential as seen from Fig.2(a).
 Thus, the ordering of the $2^-$- $1^-$  spin-doublet directly reflects the strength of the 
  $\Xi N$ interactions.  For $T=0$,   the ground (1st excited) state has {$1^- (2^-)$  
 in which dominant $\Xi N$ interaction is in the $^{11}{\rm S}_0$ ($^{13}{\rm S}_1$)} channel
 being moderately attractive (weakly attractive) in the HAL QCD potential as seen from Fig.1(a).
 Thus, the ordering of the $1^-$- $2^-$  spin-doublet directly reflects the associated  $\Xi N$ interactions too.
 An interesting feature in Fig.2(b) is that we have ``spin-doublet inversion" between 
  $T=1$ and $T=0$, which imprints  the relative 
 strength of $\Xi N$ interactions in different channels.

  {We remark here that  $\Xi N \alpha \alpha (T=0, J^{\pi}=1^-)$ is most bound 
   relative to the two-body breakup threshold with  the binding energy of  2.74 MeV (= 4.31 MeV $-$1.57 MeV)
    as seen in Fig.2(b).  This is due to the largest attraction in the  $^{11}{\rm S}_0$ $\Xi N$ channel among other 
    channels as shown in Fig.2(a).}
 Also, we find that the states in Table \ref{tab:energy}  
  have very small decay widths, {$\Gamma = 20-40$ keV.}  This is due to the fact {that the $\Xi N$-$ \Lambda \Lambda $}
  interaction appears only at a short distance in the HAL QCD potential \cite{Sasaki2020},
  so it causes only a small coupling at low energies.  This is also in accordance with the recent
  emulsion data of $\Xi$ hypernuclei \cite{Hayakawa,Yoshimoto}. 
  
{To see if $\Xi$ and $N$ interact with each other while loosely coupled to $\alpha \alpha$ core, we 
plot  the density distributions of $\Xi$, $N$ and $\alpha$ inside the ground state of $\Xi N \alpha \alpha$ 
  with $T=1$  and  $T=0$  in Fig.3(a) and Fig.3(b), respectively.
   In both cases, we find that $\Xi$ and $N$ have extended distributions outside the 
   $\alpha\alpha$ core;  root mean squire distances $\sqrt{ \langle R^2 \rangle}$ of  $\alpha$, $\Xi$ and $N$,  from the center of mass of $\alpha\alpha$ are 
   1.74 fm (1.74 fm),  3.83 fm (3.53 fm) and 3.33 fm (3.24 fm), respectively, for $T=1$ ($T=0$).
  This indicates that $\Xi N \alpha \alpha$   is one of the ideal systems to study the spin-isospin dependence of the $\Xi N$ interaction. 
  Note that    $\sqrt{ \langle R^2 \rangle}$  for $\Xi$ is  even larger than that of $N$ despite its larger mass.  This is partly because 
  $\Xi \alpha$ attraction is  weak.}

\

{\it Summary and concluding remarks.$-$} 
We explored a new possibility to study the two-body $\Xi N$ interaction,  especially its spin-isospin dependence,
 through the energy levels of   the $\Xi N\alpha$ and $\Xi N\alpha \alpha$ systems where $\alpha$ plays a role to
  attract $ \Xi N$ pair without changing its spin-isospin structure.
 By using the Gaussian expansion method (GEM) for three- and four-body cluster model
  and  the $\Xi N$ potential obtained from the state-of-the-art 
 lattice QCD calculations, we found that
   $\Xi N\alpha \alpha$ has 
   {spin-doublet  bound states, $J^{\pi}=1^-, 2^-$ in each isospin channel, }
   while it is unlikely that $\Xi N\alpha$ has a bound state.

 The ordering of the bound state levels  of  $\Xi N\alpha \alpha$ has a characteristic structure 
 associated with the spin-isospin dependence of the $\Xi N$ interaction (Fig.2(a)):
 We found the inversion of the $1^-$-$2^-$  spin-doublet between $T=1$ and $T=0$ (Fig.2(b)).
Also,  the largest $\Xi N$ attraction in the $^{11}{\rm S}_0$ channel is reflected in the 
  largest binding energy of the $(T, J^{\pi})=(0, 1^-)$ state relative to the two-body breakup threshold.
  
 These level structures of $\Xi N \alpha \alpha$ bound states can be studied
experimentally by  producing  $\Xi N\alpha \alpha$ in $T=1$ and $T=0$ states
 through   $(K^-, K^+)$ and $(K^-,K^0)$ reactions on the  $^{10}$B  target, respectively.
 If the level ordering between the  $1^-$ and $2^-$ states is determined by the reaction cross section, 
  information on the relative strengths of the   $\Xi N$ interactions  in $^{33}{\rm S}_1$ and $^{31}{\rm S}_0$, $^{13}{\rm S}_1$ and $^{11}{\rm S}_0$ channels
    can be extracted.  
  At J-PARC facility, such experiments could be pursued after the planned experiment to
   produce $^{12}_\Xi$Be and $^7_{\Xi}$H using $^{12}$C and $^7$Li target by the $(K^-,K^+)$ reaction \cite{fujioka}.
{For an accurate comparison with future experimental data, theoretical calculations that fully account for isospin symmetry breaking are 
 needed.}
 
\
 
\begin{acknowledgments}
{\it Acknowledgments.$-$}
 The authors thank Makoto Oka for illuminating discussions.
The work was supported in part by JSPS KAKENHI Grant Numbers JP18H05236, {JP19K03879} and JP20H00155,
Grant-In-Aid for scientific research on Innovative Areas,
 JP18H05407,  the National Science Foundation under Grant No. NSF PHY-1748958,
and Pioneering research project 'RIKEN Evolution of Matter in the
Universe Program'.
 This work was partially supported by 
``Program for Promoting Researches on the Supercomputer Fugaku''
 (Simulation for basic science: from fundamental laws of particles to creation of nuclei)
  and Joint Institute for Computational Fundamental Science (JICFuS).
\end{acknowledgments}

\end{document}